\documentclass[conference]{IEEEtran}

\IEEEoverridecommandlockouts
\usepackage{cite}
\usepackage{amsmath,amssymb,amsfonts}
\usepackage{algorithmic}
\usepackage{graphicx}
\usepackage{textcomp}
\usepackage{xcolor}
\def\BibTeX{{\rm B\kern-.05em{\sc i\kern-.025em b}\kern-.08em
    T\kern-.1667em\lower.7ex\hbox{E}\kern-.125emX}}
\begin{document}

\title{Accelerating Deep Neural Networks for Real-time Data Selection for High-resolution Imaging Particle Detectors\\
%\thanks{\textcolor{red}{funding acknowledgements here}}
}

\newif\ifdraft
\draftfalse
%\drafttrue

\author{
\IEEEauthorblockN{Yeon-jae Jwa}
\IEEEauthorblockA{\textit{Dept. of Physics} \\
\textit{Columbia University}\\
New York, NY, USA \\
yj2429@nevis.columbia.edu}
\and
\IEEEauthorblockN{Giuseppe Di Guglielmo}
\IEEEauthorblockA{\textit{Dept. of Computer Science} \\
\textit{Columbia University}\\
New York, NY, USA \\
giuseppe@cs.columbia.edu}
\and
\IEEEauthorblockN{Luca P. Carloni}
\IEEEauthorblockA{\textit{Dept. of Computer Science} \\
\textit{Columbia University}\\
New York, NY, USA \\
luca@cs.columbia.edu}
\and
\IEEEauthorblockN{Georgia Karagiorgi}
\IEEEauthorblockA{\textit{Dept. of Physics} \\
\textit{Columbia University}\\
New York, NY, USA \\
georgia@nevis.columbia.edu}
}

\maketitle

\begin{abstract}
This paper presents the custom implementation, optimization, and performance evaluation of convolutional neural networks on field programmable gate arrays, for the purposes of accelerating deep neural network inference on large, two-dimensional image inputs. The targeted application is that of data selection for high-resolution particle imaging detectors, and in particular liquid argon time projection chamber detectors, such as that employed by the future Deep Underground Neutrino Experiment. We motivate this particular application based on the excellent performance of deep neural networks on classifying simulated raw data from the DUNE LArTPC, combined with the need for power-efficient data processing in the case of remote, long-term, and limited-access operating detector conditions. 
\end{abstract}


\section{Introduction}
\label{sec1}

Liquid Argon Time Projection Chambers (LArTPCs) represent a particle detector technology that has been widely adopted in the field of high energy physics. Over the last two decades, LArTPCs have been increasingly used for studying neutrino-argon interactions with high calorimetric (energy) and spatial resolution. LArTPCs are already in use for a number of detectors; the most recent of these detectors, MicroBooNE \cite{ubdet} and ProtoDUNE \cite{protodune}, represent a significant R\&D effort which is underway to scale up the LArTPC detector technology by up to two orders of magnitude in physical detector size. This phasing approach is necessary in order to realize the future Deep Underground Neutrino Experiment (DUNE) \cite{duneidr,dunetdr}, which will feature the largest LArTPC detector to be ever constructed and operated at a deep underground location in Lead, South Dakota, in the United States, starting in $\sim$2025.

LArTPCs, including DUNE, work by imaging particle tracks and other signatures imprinted in a large, uniform detector volume by particles produced in neutrino or other rare physics interactions. Different interactions yield distinct image topologies that are identifiable and differentiatable by their spatial extent, shape, and pixel intensity, when viewed as two-dimensional projections of a three-dimensional detector region. Furthermore, the format of the detector-generated raw data represents exactly two-dimensional projections of the activity inside the detector; as such, a potentially advantageous solution for real-time data processing and data selection (triggering) on interesting detector activity is image analysis with hardware-accelerated Deep Neural Networks (DNNs). 

DNNs are already being applied successfully for the offline analysis of data recorded by existing high energy physics experiments \cite{Radovic:2018dip}, including operating LArTPCs. In the case of the latter, MicroBooNE is pioneering the use of deep learning for neutrino physics analyses (see, e.g., \cite{uboonecnn,ubpid}), and similar DNN-based methodologies have now been adopted for several analyses planned with the future DUNE experiment \cite{dunetdr}. Machine learning approaches to LArTPC data analysis are gaining increasing traction (see, e.g.~\cite{sparsecnn}); meanwhile, new techniques are continually being considered to improve data processing latency and resource requirements, with promising results \cite{sze}.

At the same time, the success of DNNs more generally has motivated the research and development of many specialized system architectures and accelerators both in academia and in industry. %For instance, the Intel Knights Mill processor implements special vector instructions for deep learning and the NVIDIA PASCAL GP100 GPU features 16-bit floating-point (FP16) arithmetic support to perform two FP16 operations simultaneously on a single-precision core. 
An excellent overview of the challenges of accelerating DNNs in hardware and a comprehensive survey of many techniques and frameworks that have been proposed so far in the literature is provided in \cite{falsafi}. In terms of implementation, DNN frameworks mainly target CPUs and GPUs. In particular, GPUs offer high computational density and high level of programmability; this simplifies the interface with operating systems while providing access to powerful computational platforms for data-parallel algorithms and dense floating-point operations. GPU performance, however, comes with high power dissipation, making a GPU-based solution unsustainable for many high-performance embedded systems that require major power efficiency. Thanks to their hardware reconfigurability, Field Programmable Gate Arrays (FPGAs) are a valid alternative solution as power-aware platforms for DNN acceleration \cite{falsafi}. In addition to hardware developments, frameworks such as Caffe \cite{jia} and Tensorflow \cite{abadi} allow a much larger user base for modern DNNs.

In this paper, we investigate
the viability of DNN implementations in a variety of architecture systems, including GPU for online data processing, and FPGA or mixed FPGA-CPU architecture systems for real-time data processing, both for the purposes of data selection (triggering) for a high-resolution and high-rate imaging detector. The application we specifically target is that of DUNE, which involves real-time streaming of data rates of the order of tens of terabits per second. The proposed data selection schemes, however, may be applicable to any LArTPC, sharing the same technology as DUNE, and particularly viable for smaller-scale ones. We note that the application of machine learning algorithms for triggering purposes has been considered for other types of particle detectors (see, e.g.~\cite{Duarte:2019fta}). However, the application proposed here for LArTPCs is a new effort, and it deals with a unique set of challenges: specifically, LArTPC triggering is governed by a much larger input (image) size, but also benefits from relaxed latency constraints due to a much slower detector response than other types of particle detectors. The targeted DUNE application and DUNE detector design are presented in Sec.~\ref{sec2}. 

To motivate the application of DNNs for DUNE data selection purposes, we train and investigate the performance of a number of DNNs on simulated LArTPC raw data images. Results obtained on GPUs are presented in Sec.~\ref{sec3}, and demonstrate high efficiency in selecting rare physics interactions of interest, while maintaining a sufficiently low selection rate from background interactions and detector noise. Latency and power dissipation considerations, however, motivate the investigation of inference on FPGA or mixed FPGA-CPU systems, which have been shown to achieve significant speedup \cite{coussy}. 
%To motivate DNN application for DUNE data selection purposes, one must demonstrate high efficiency in selecting rare physics interactions of interest, while maintaining a sufficiently low selection rate from background interactions and detector noise. To do so, we train a number of DNNs with simulated LArTPC images in GPUs, and test the trained network models in order to investigate signal efficiency and background rejection; results are provided in Sec.~\ref{sec3}. In Sec.~\ref{sec4}, we further investigate and contrast DNN scalability and applicability toward DUNE for a number of DNNs, in terms of latency and power dissipation. We find that CNN inference latency on GPUs is currently a limiting factor for some implementations, while power utilization is an additional concern. On the other hand, inference on FPGA or mixed FPGA-CPU systems has been shown to achieve significant speedup \cite{coussy}. 
As such, in Sec.~\ref{sec4}, we present several contributions for designing hardware acceleration of Convolutional Neural Network (CNN) inference algorithms on resource-constrained platforms like FPGAs. By using a customizable and efficient hardware accelerator design for the various layers, we show that the flexibility of the accelerator design together with the possibility of leveraging the knobs provided by High Level Synthesis (HLS) tools enable the design of high-performance accelerators that can greatly benefit the deployment of DNN models. Finally, in Sec.~\ref{sec5}, we identify DNNs which would satisfy DUNE physics and latency requirements, considering also resource utilization on an FPGA with specifications that might be suitable for DUNE readout.

\section{Application Use Case: Deep Underground Neutrino Experiment}
\label{sec2}

DUNE is an international particle physics experiment that aims to study neutrinos and their oscillation patterns with unprecedented sensitivity as well as search for other rare particle interaction signatures that will inform our understanding of nature at the most fundamental level. In particular, DUNE measurements aim to elucidate the underlying mechanism responsible for the prevalence of matter over antimatter in our observable universe. To accomplish these physics goals, the DUNE far detector will employ four LArTPC modules, each holding 10 kilotons of liquid argon in total fiducial detector mass, and will operate for more than a decade in a deep underground location at Sanford Labs, in Lead, South Dakota, beginning in the middle of the next decade. 

To study neutrino oscillations, DUNE must detect interactions of neutrinos from a high-intensity pulsed beam from Fermi National Accelerator Lab, in Batavia, Illinois. Selecting and recording these interactions is straightforward since they are all expected to arrive only during a relatively short time dictated by the beam pulse structure; the latter is precisely known due to external beam timing signals informing the trigger decision. To study other rare, off-beam events such as proton decay events, neutron-antineutron oscillation events, and interactions of neutrinos from galactic supernova bursts, however, DUNE must continually process its data in order to make a data-driven decision to select and record these signatures. This is because these signatures are random in nature, and no prompt external timing signal is available to independently inform the data selection decision. The expected rate of rare off-beam events, and other off-beam interactions of interest, in DUNE is provided in Tab.~\ref{tab1}. 

\begin{table}[t]
\caption{Expected rates of rare off-beam events and other off-beam signatures in a 10~kton (fiducial mass) DUNE far detector module \cite{dunetdr}. }
\begin{center}
\begin{tabular}{|l|l|l|}
\hline
\textbf{Interaction Type}&\textbf{Event Type}&\textbf{Expected Rate} \\
\hline
\multicolumn{3}{|l|}{\textbf{Rare off-beam events}}\\
\hline
Proton decay & High Energy (HE) & $<1$ / year\\ 
\hline
Neutron-antineutron oscillation & High Energy (HE) & $<1$ / year\\
\hline
Galactic supernova burst$^{\mathrm{a}}$ & Low Energy (LE) & $<1$ / year\\
\hline
\multicolumn{3}{|l|}{\textbf{Other off-beam events}} \\
\hline
Atmospheric neutrinos & High Energy (HE) & 1200 / year\\ 
\hline
Cosmic ray muons & High Energy (HE) & 1.3$\times10^6$ / year\\ 
\hline
\end{tabular}
\label{tab1}
\end{center}
$^{\mathrm{a}}$A galactic supernova burst is expected at a rate of roughly once per century. The latest galactic supernova burst was observed in 1604 \cite{snb}.
\end{table}

The DUNE system responsible for data selection must, in the end, only allow for effectively 30 petabytes of data per year to be diverted to permanent storage offline \cite{dunetdr}. As such, given the multiple tens of terabits per second raw data rate of the DUNE far detector, a factor of $10^{4}$ data reduction must effectively be achieved by the system, without compromising efficiency for selecting rare events of interest. Generally, a trigger efficiency of $>$99\% is required for high energy events, including atmospheric neutrino interactions, proton decay events, neutron-antineutron oscillation events, and cosmic ray muon events in the detector. Similar trigger efficiency is also required for selecting aggregates of multiple low energy supernova neutrino interactions that are expected to occur in case of a galactic supernova burst. In that case, the trigger efficiency requirement on any individual supernova neutrino interaction can be relaxed, and a multiplicity condition can be used to boost efficiency for coincident interactions\footnote{In the case of a supernova at the edge of our galaxy, for example, approximately 50 supernova neutrino interactions are expected over the span of ten seconds in each DUNE 10~kton module.}.  

The main challenge, specific to the supernova burst trigger, is that individual supernova events are characterized by low energy deposition in the detector; as such, their observable signature is similar to that of intrinsic radiological backgrounds and electronics noise in the detector, which are the dominant contributor to observable signals in the DUNE data. Consequently, in order to achieve the desired data reduction factor, significant noise and radiological background rejection is needed.

Two distinct detector designs are in development for the DUNE far detector modules. We restrict the discussion and studies presented in this paper to the so-called ``single phase'' LArTPC module technology, described in the following subsection, following \cite{dunetdr}. However, both the single phase and ``dual phase'' technology operate on high-resolution imaging principles; we therefore expect that comparable challenges, solutions, and performance would be achievable for the ``dual phase'' technology for DUNE as well.

\subsection{DUNE Single Phase Detector Design}
In the case of the DUNE far detector ``single phase'' design, each DUNE 10 kton far detector module is segmented into 150 individual ``cells'' (rectangular volumes) of liquid argon, which are imaged by sensor-wire arrays, called an Anode Plane Arrays (APAs). An APA is positioned in the middle of each cell, and it consists of multiple planes of parallel wires oriented in three distinct directions relative to the vertical direction. The wires sense ionization charge (electrons) liberated by charged particles along the charged particles' paths as they traverse the liquid argon volume enclosed in the cell; the ionization charge drifts toward the wire planes under the influence of a strong, uniform electric field applied across each cell, on either side of the APA. Given the arrival time of the ionization charge, relative to the time of the interaction (identified and recorded by detecting the prompt scintillation light produced at the time of the interaction, using a dedicated photon detection system), the drift coordinate of the event can be reconstructed. The ionization signals recorded as a function of wire number across each wire plane, and as a function of time, can then be mapped into a two-dimensional projected view of the cell, for a given time; this makes it possible to reconstruct a three-dimensional view of any interaction inside a cell by matching signals across the three stereoscopic views (one per plane). 

The studies in this paper involve only signals from vertically oriented wire planes. One such plane exists on each side of the APA, and makes up a so called charge ``collection'' wire plane. Due to the electric field configuration and readout electronics response, recorded signals on collection wires are unipolar; as such, their amplitudes and integrals, in particular, correlate highly with the amount of ionization charge arriving at each wire. We refer to channel vs.~time data which spans the equivalent of a collection plane times drift time (drift length on one side of the APA divided by wire signal sampling rate) as an ``APA-frame''.  For the DUNE APA cell physical dimensions and nominal electric field configuration, the APA-frame drift time corresponds to 2.25~ms. 

\begin{figure*}[h!]
\centerline{\includegraphics[width=\textwidth, trim=0 2.82in 0 0.22in, clip]{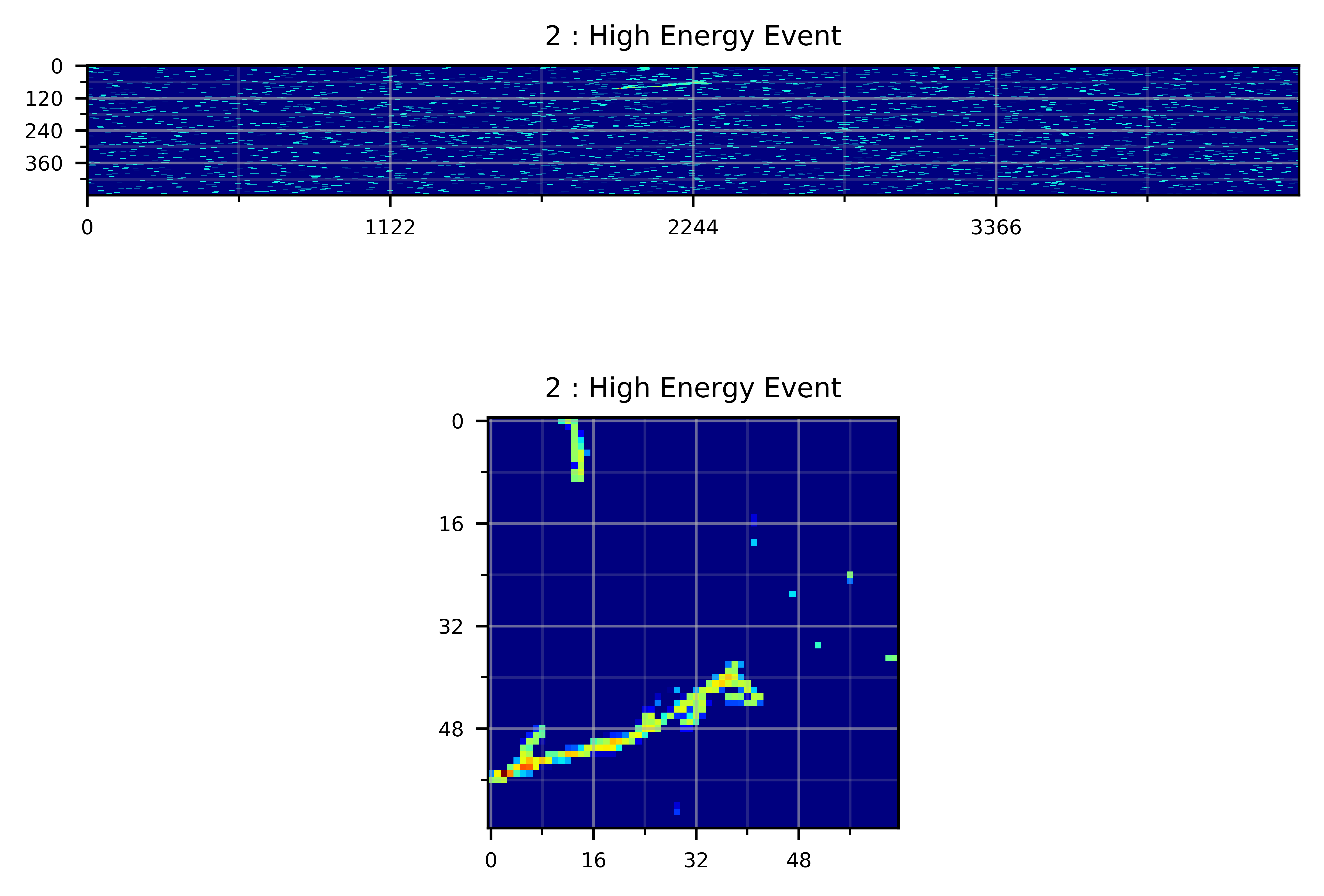}}
\ifdraft
\else 
\centerline{\includegraphics[width=\textwidth, trim=0 2.82in 0 0.22in, clip]{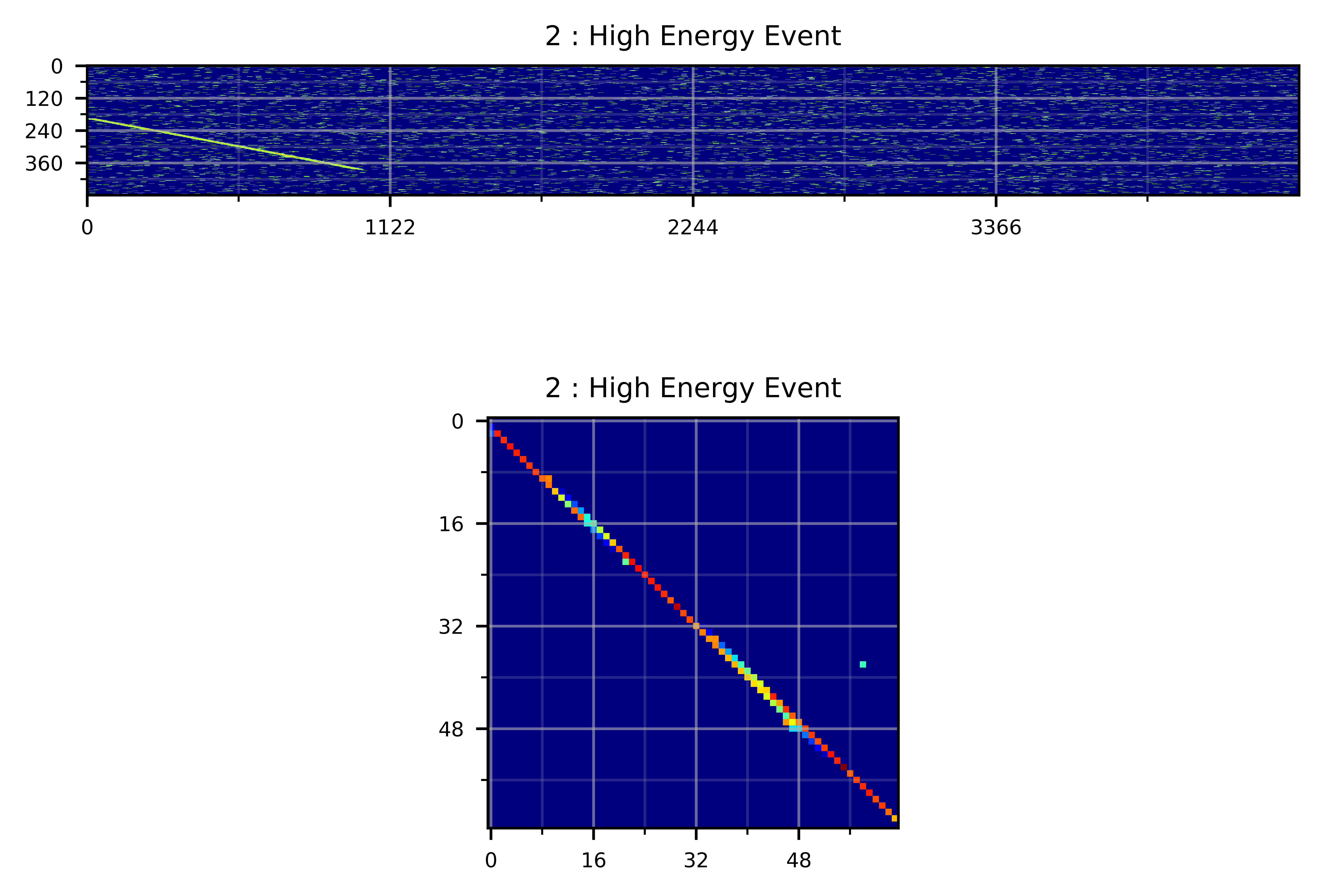}}
\centerline{\includegraphics[width=\textwidth, trim=0 2.82in 0 0.22in, clip]{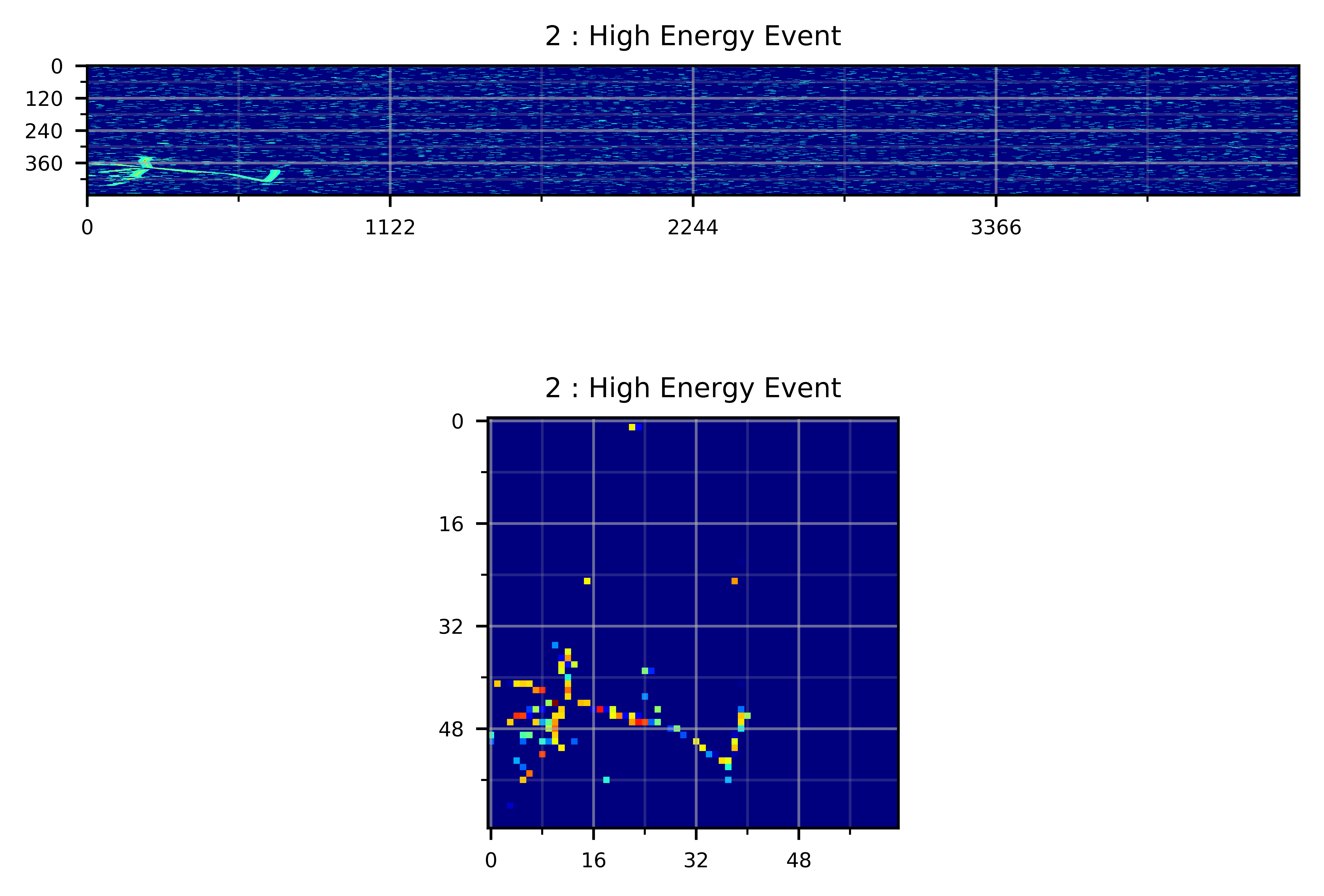}}
\centerline{\includegraphics[width=\textwidth, trim=0 2.82in 0 0.22in, clip]{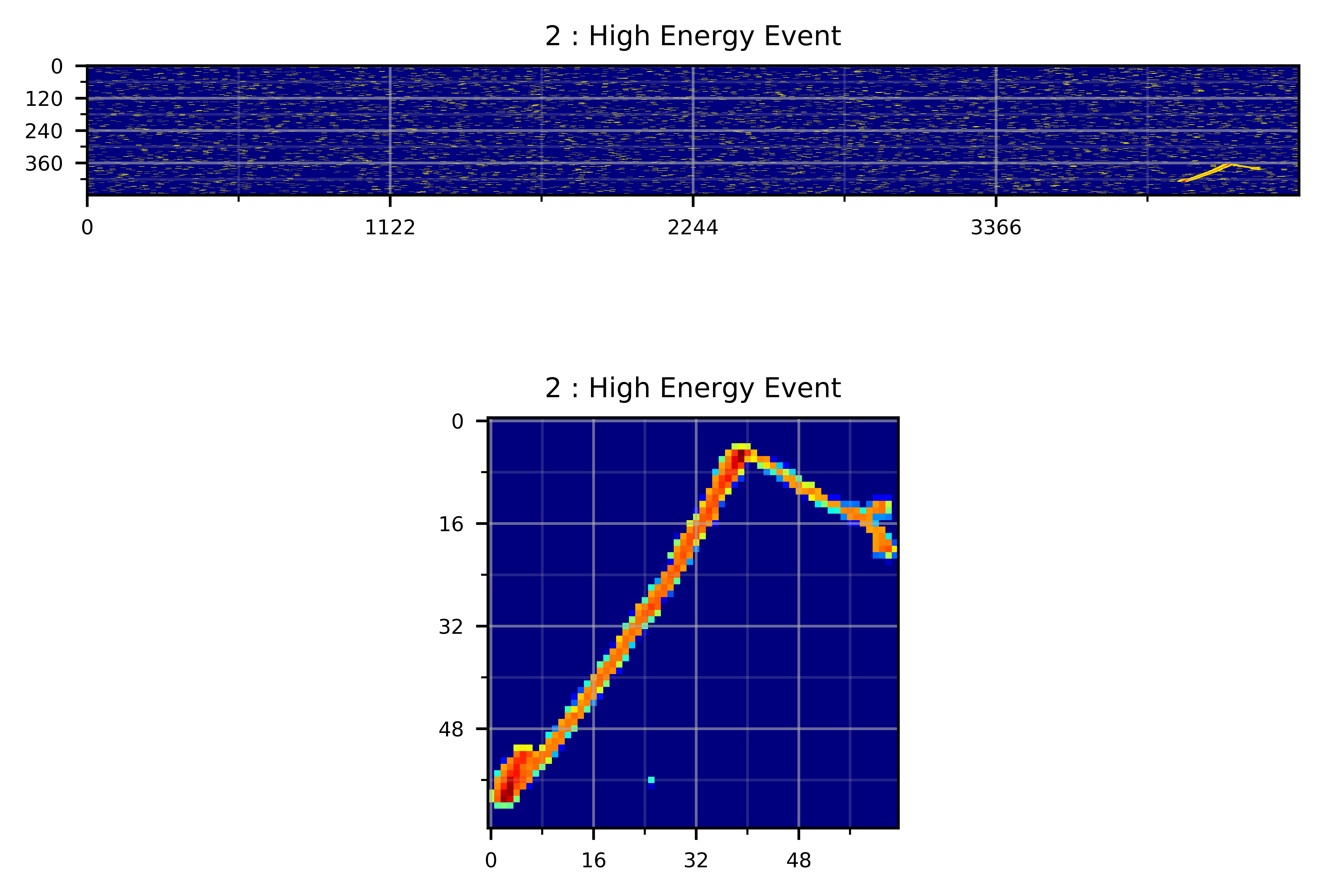}}
\centerline{\includegraphics[width=\textwidth, trim=0 2.82in 0 0.22in, clip]{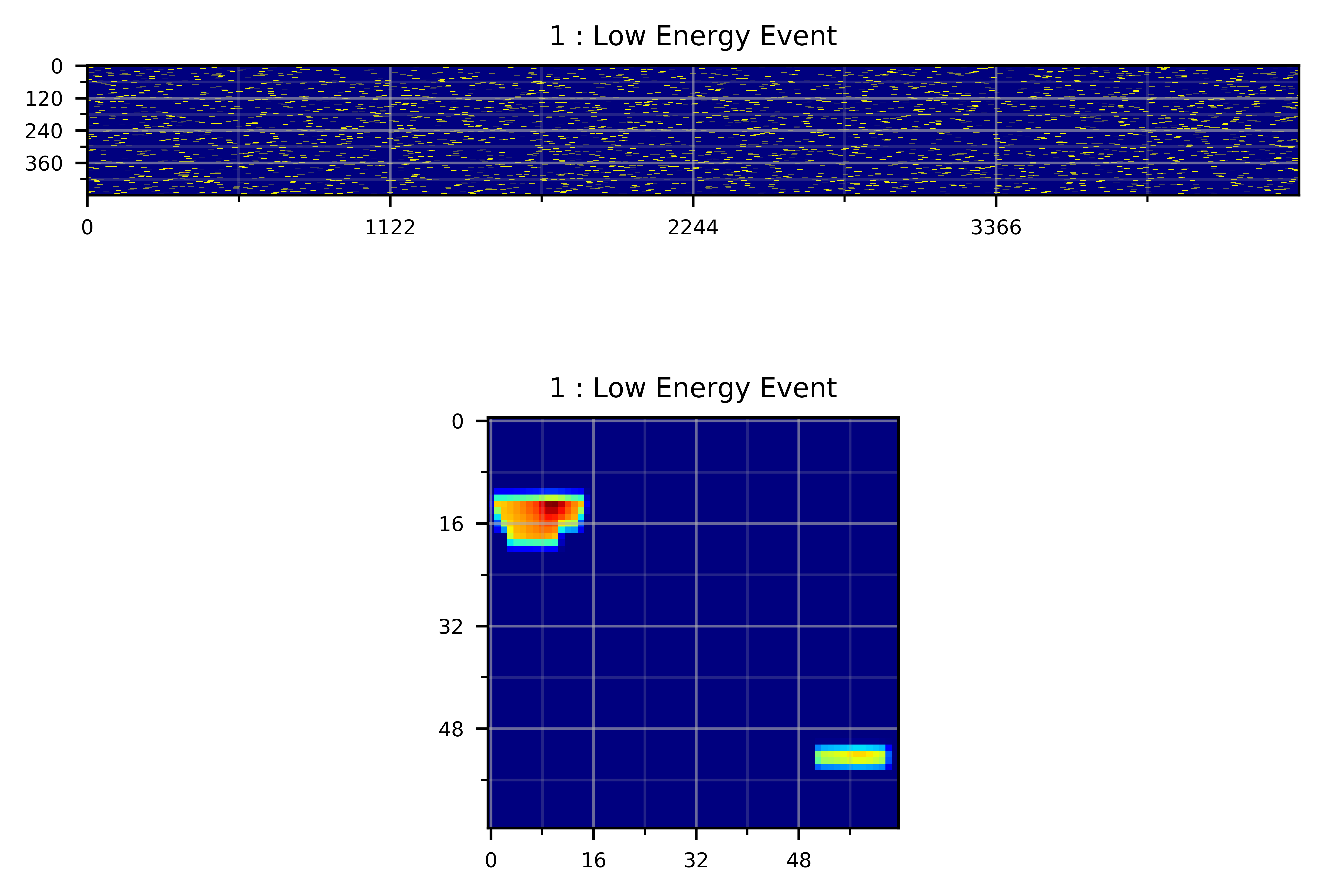}}
\centerline{\includegraphics[width=\textwidth, trim=0 2.82in 0 0.22in, clip]{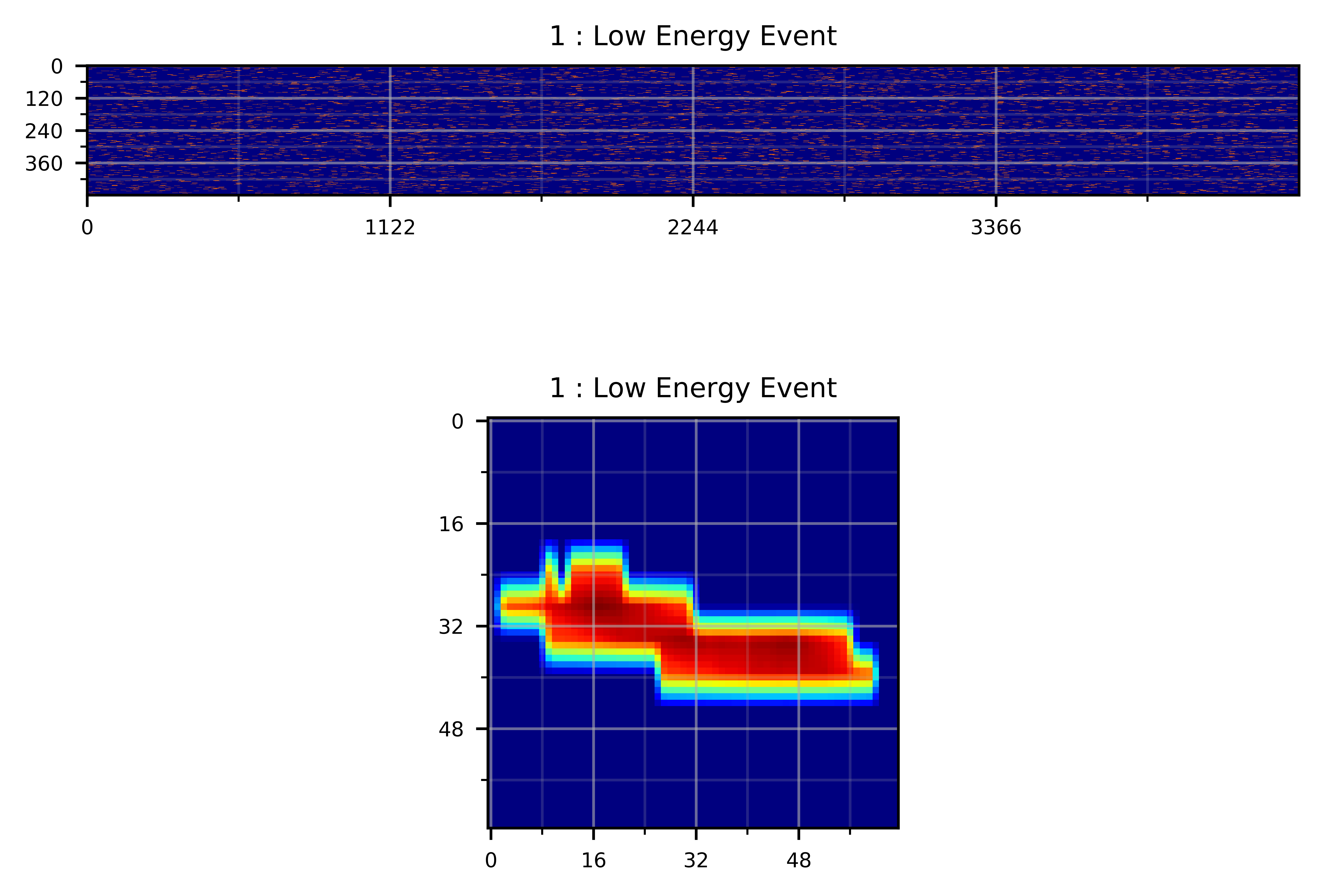}}
\centerline{\includegraphics[width=\textwidth, trim=0 2.82in 0 0.22in, clip]{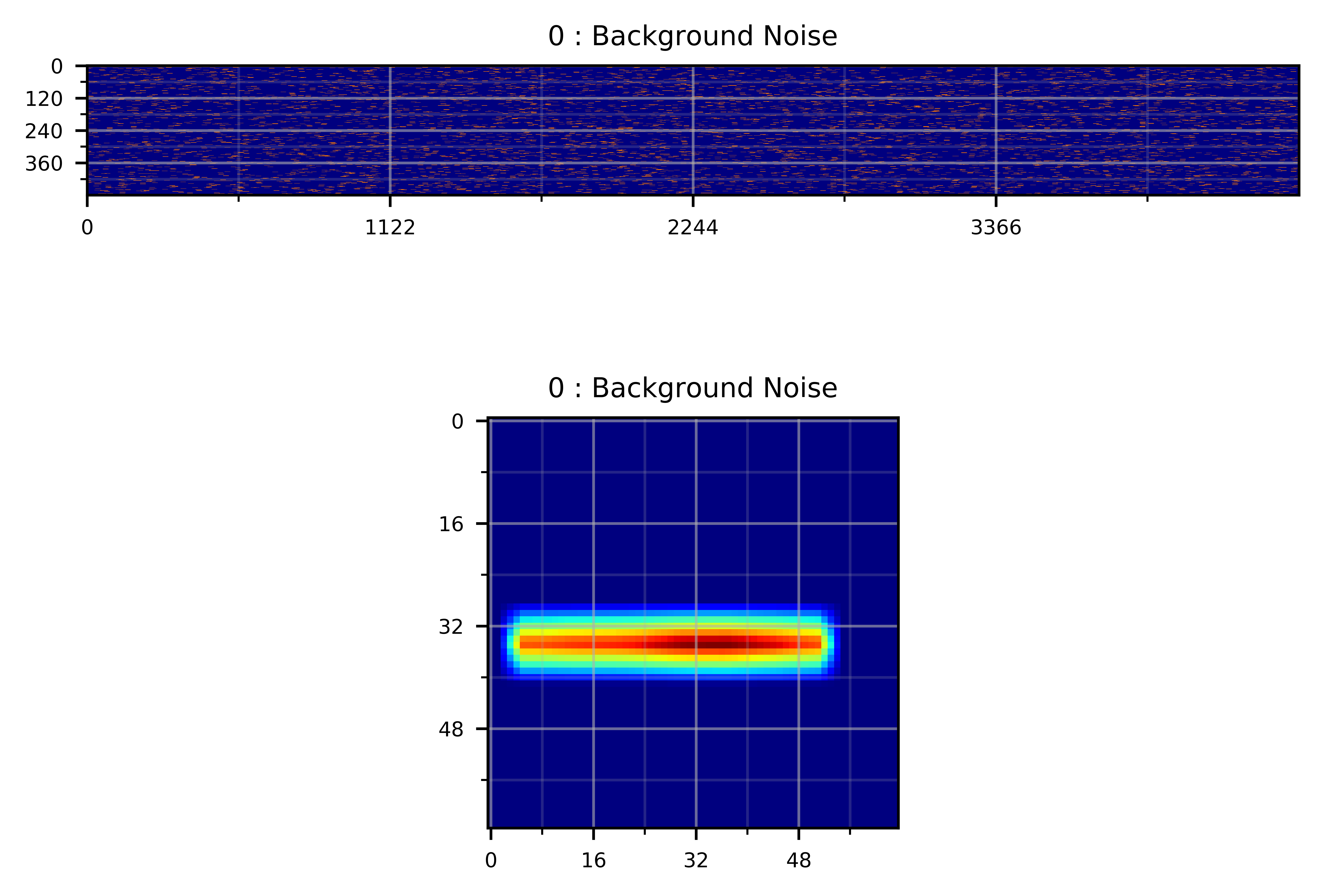}}
\centerline{\includegraphics[width=\textwidth, trim=0 2.65in 0 0.22in, clip]{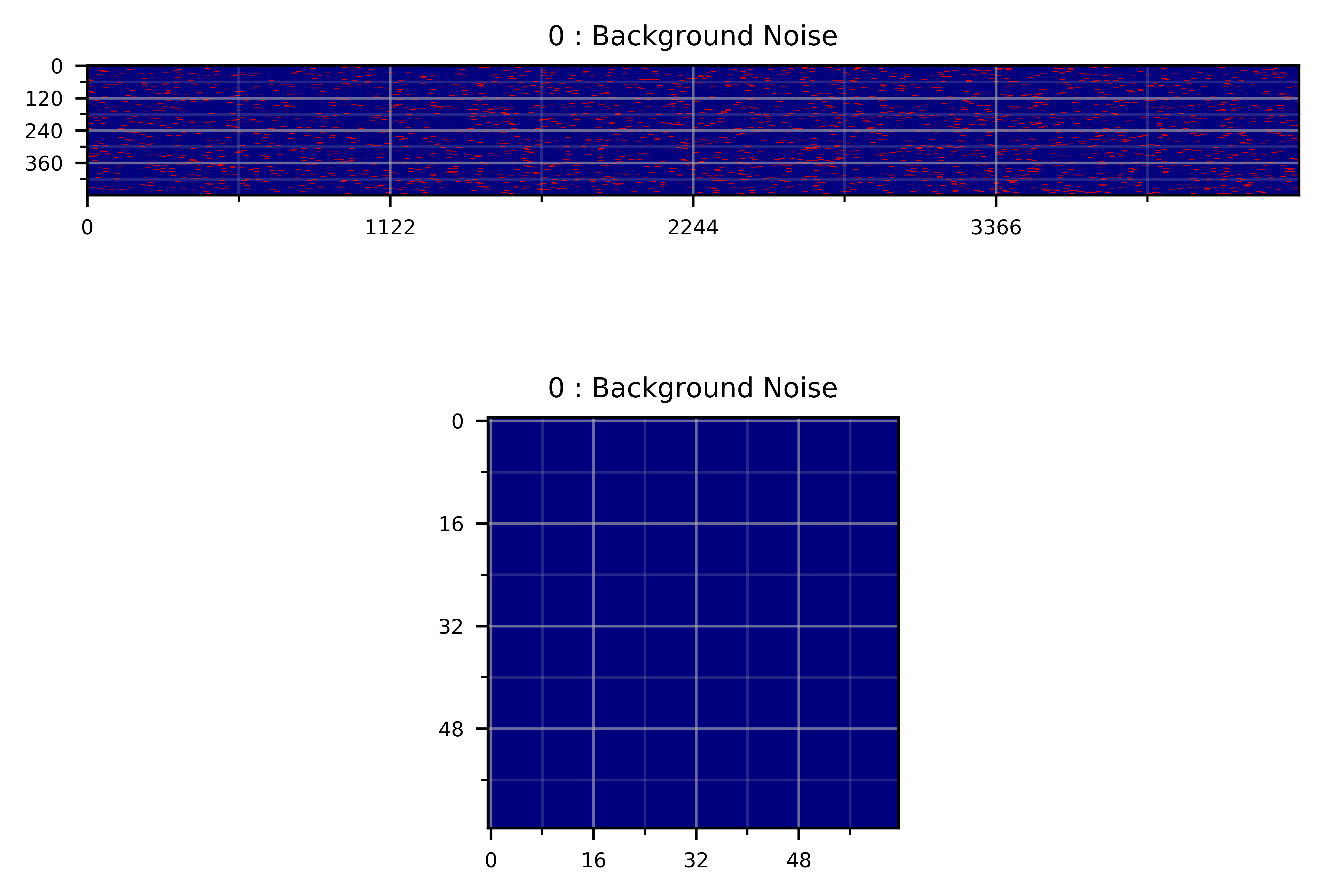}}
\fi
\caption{Simulated APA-frames, representative of three main types of signatures of interest. The top four frames correspond to high energy events; the lower four frames correspond to low energy supernova neutrino events (first two) and empty events including only background noise (bottom two). APA-frames are defined according to the APA drift volume in which the interaction originates. The $y$ axis of each frame corresponds to collection plane channel number; the $x$ axis corresponds to time tick (2~MHz) across a full 2.25~ms readout.}
\label{fig1}
\end{figure*}

Simulations of APA-frames representative of several topologies of interest, from Tab.~\ref{tab1}, are show in Fig.~\ref{fig1}. APA-frames are simulated using the LArSoft framework \cite{larsoft6, larsoft7} and DUNE Monte Carlo generation tools \cite{dunetpc}. DUNE Monte Carlo generation configuration parameters are set to the dunetpc v07\_13\_00 default values, except for the electronics noise RMS levels, which are artificially enhanced for conservatism; specifically, in our simulations, we increase the collection plane electronics noise RMS by 40\% relative to the default value. All APA-frames with topologies of interest also include default radiological background and electronics noise. %We found no significant changes could bring differences in the generation regarding the objects we use in this study between two dunetpc releases \cite{breakingchanges}.  

\subsection{DUNE Data Acquisition System Design}

DUNE will have to operate continually, for more than a decade, streaming data out of its LArTPC detectors
at a total rate of multiple tens of terabits per second. For reference, MicroBooNE \cite{ubdet} and ProtoDUNE \cite{protodune}, the two largest currently operating LArTPCs, stream images continually at a data rate of greater than 260 and 490 gigabits per second, respectively. %Example images recorded by the MicroBooNE detector, with and without data reduction applied, are shown in Fig.~\ref{xx}.  
Unlike DUNE, these experiments do not have a rare event search physics scope. Data reduction for these detectors is therefore achieved through a combination of external trigger signals informing when to record a small subset of that data, and additional real-time compression, filtering and/or zero-suppression carried out in FPGA and/or CPU (see, e.g.~\cite{ubdet,ub-pub-note}). 

Differently from these detectors, the DUNE detector must be capable of processing its data in real time, or, in an online fashion, in order to make data-driven decisions to record what might be rare physics events. DUNE's data acquisition system (DAQ), and in particular its data selection (sub)system, must do so with negligible dead-time, to maximize the detector's physics sensitivity
to rare signatures. An additional constraint is power distribution limitations at the (underground) detector
site. Specifically, the DUNE far detector DAQ is limited to 500~kVA of power underground, or 125~kVA per 10 kton module, plus an additional 50~kVA of power available on the surface for back-end DAQ \cite{dunetdr}.

The baseline DUNE DAQ design is documented in detail in \cite{dunetdr}. It employs a multi-level data selection system. First, a low-level data selection decision is achieved on a combination of CPU and FPGA resources. %It is estimated that a single FPGA chip should be responsible for processing .
%a module-level data selection decision is achieved 
This level of data selection is executed independently on a per-APA basis, while the second level, to first order, aggregates low-level information from all APAs in a single module to make a module-level trigger decision. The module-level trigger decision is executed on CPU resources, and its latency is limited to a few seconds. When formed, a module-level trigger decision instructs the readout of several milliseconds worth of continuous data from all 150 APAs in the module, or, in the case of a supernova burst trigger decision, 100 seconds worth of continuous data from all 150 APAs. Non-supernova burst trigger decision rates of up to O(1)~Hz are possible, while supernova burst trigger decision rates are limited to one per month. These upper limits on trigger rates include fake triggers on accidental backgrounds and noise; therefore, background noise considerations are especially important in the case of supernova burst triggers. Additional data down-selection can be achieved by the use of a high-level %filter farm (GPU resources),
filter farm, 
which is envisioned to employ data selection techniques similar to the ones presented in this work. 

\section{DNN-based LArTPC Data Selection}
\label{sec3}

To motivate DNN-based LArTPC data selection, we have studied a number of DNNs in terms of their performance on classifying simulated DUNE far detector single phase APA-frames. We have considered a multi-class data classification scheme, where the different classes represent different types of off-beam physics events of interest that can occur in the DUNE far detector, as well as non-physics events (intrinsic to the detector materials radiological backgrounds and electronics noise backgrounds). 

The methodology we followed assumes that a two-level data selection system is used to (1) first generate a low-level data selection decision, specifically the classification of APA-frames according to their content with the use of a DNN, and to (2) subsequently process those decisions further in order to make a module-level data selection decision. More specifically, the
module-level data selection stage keeps track of information\footnote{Spatial coordinate, type of interaction, etc.} from APA-frames that have been tagged as a
certain type of interaction over the entire 10~kton detector module, over a given time interval. 
In this way, for example, a supernova burst trigger decision can be generated at the module level if multiple APA-frames are tagged
by the ``low-level'' trigger as containing supernova neutrino interactions over a short amount of
time (typically on the order of seconds). Our studies focus particularly on the low-level stage of processing.

The APA-frames stream continually from each DUNE detector module, at a rate of 200 frames (one for each drift volume) per 2.25~ms. Each frame is 480 channels wide by (2.25~ms)$\times$(2~MHz)$=$4500 samples\footnote{More specifically, 4488 samples are used for simulation purposes.} wide, corresponding to a total of 4.15 megapixels, with 12-bit color resolution. Because of the large APA-frame size (3.2~GB), significant down-sizing is necessary in order to fit APA-frames into image sizes typically processed by DNNs. Down-sizing is also applied in anticipation of the limited resources available on FPGAs that the DUNE far detector data selection system will employ for low-level data selection \cite{dunetdr}, which we consider to be a candidate hardware platform for DNN deployment. 

\begin{figure}[t]
\includegraphics[width=0.5\textwidth]{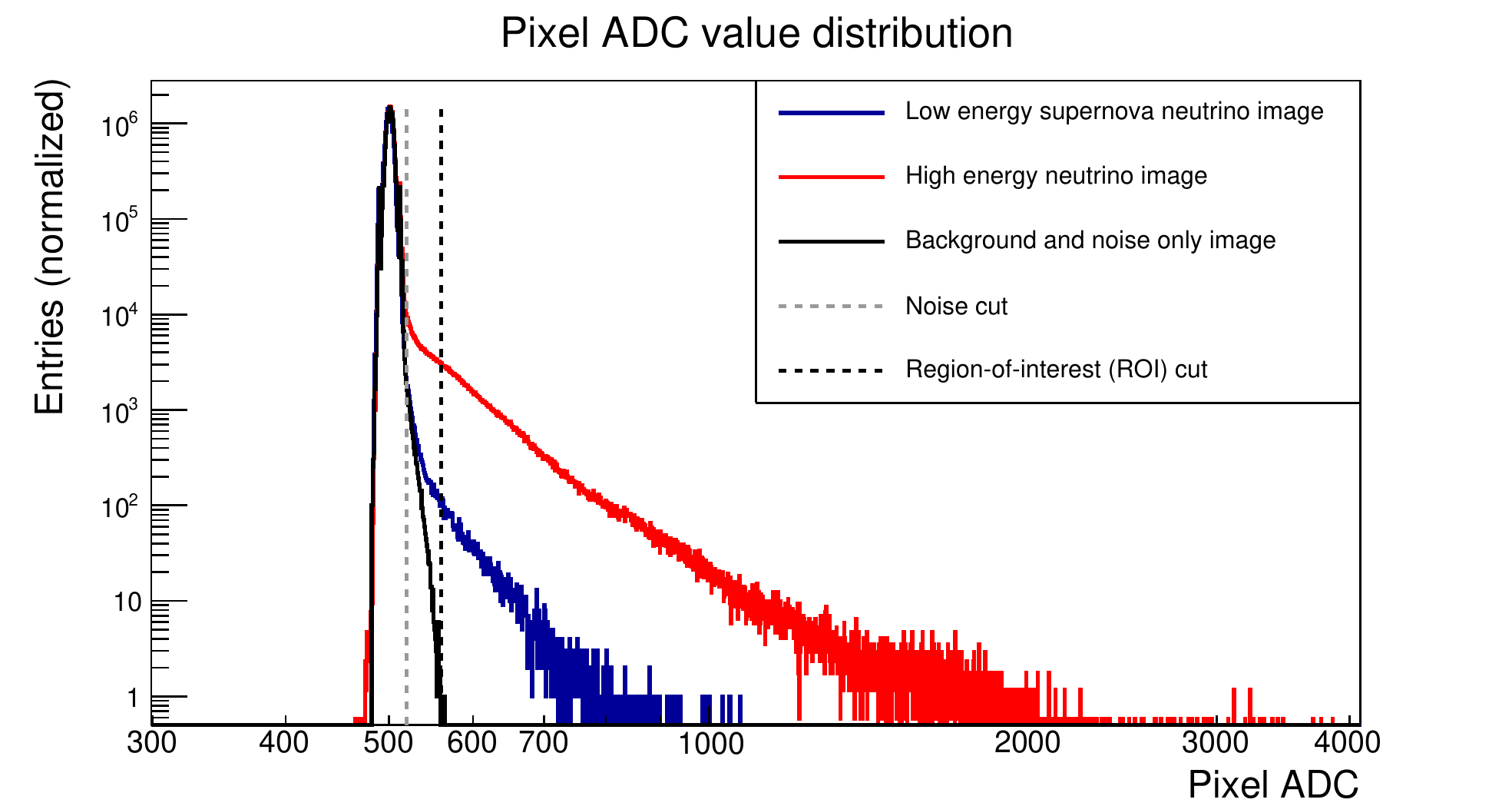}
\caption{Pixel ADC distributions of frame images for the three classes in consideration: background noise (black), low energy neutrino (blue), and high energy neutrino (red) images. The pixel ADC values range from 0 to 4095 (12-bit ADC). The distributions are absolutely normalized to 100 images with 480$\times$4488 pixels each. The background noise distribution peaks below 520 ADC for all frames. The dashed vertical lines indicate cuts that were used in pre-processing input images for the networks, in order to de-noise the raw images and to select regions with candidate physics interactions. Based on these distributions, a noise removal cut (indicated by the dashed gray line) and an ROI cut (indicated by the dashed black line) was set to 520 and 560~ADC, respectively.}
\label{fig2}
\end{figure}

\begin{figure}[t]
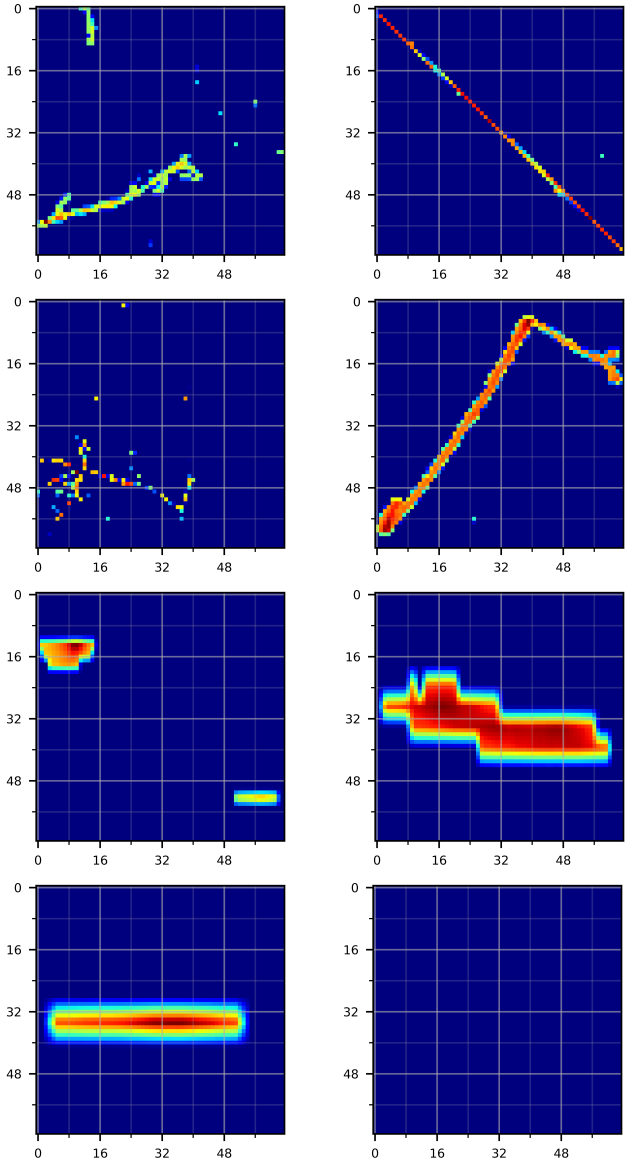

\includegraphics[width=0.24\textwidth, trim=1.6in 0 1.6in 1.68in, clip]{figures/HE_image-0-10.png}
\includegraphics[width=0.24\textwidth, trim=1.6in 0 1.6in 1.68in, clip]{figures/HE_muon_image-0-1.png}
\ifdraft
\else
\includegraphics[width=0.24\textwidth, trim=1.6in 0 1.6in 1.68in, clip]{figures/HE_nnbar_image-0-45.png}
\includegraphics[width=0.24\textwidth, trim=1.6in 0 1.6in 1.68in, clip]{figures/HE_other_image-0-78.png}
\includegraphics[width=0.24\textwidth, trim=1.6in 0 1.6in 1.68in, clip]{figures/LE_SN1_image-0-6.png}
\includegraphics[width=0.24\textwidth, trim=1.6in 0 1.6in 1.68in, clip]{figures/LE_SN2_image-0-20.png}
\includegraphics[width=0.24\textwidth, trim=1.6in 0 1.6in 1.68in, clip]{figures/N_image-0-155.png}
\includegraphics[width=0.24\textwidth, trim=1.6in 0 1.6in 1.68in, clip]{figures/N_image-0-156.png}
\fi
\caption{ROIs extracted using Method 2 for the simulated frames shown in Fig.~\ref{fig1}.  The $y$ axis represents channel space; the $x$ axis represents time space. The top four panels correspond to high energy interactions; the subsequent two correspond to low energy interactions; the bottom two correspond to background noise (typically empty frames, after noise removal, or noise artifacts). Noise removal is achieved by zero-suppressing pixels with ADC values below 520 ADC; an ROI is defined by first finding the smallest contiguous rectangular region in a frame that contains at least one pixel value exceeding 560 ADC, padded by five (5) additional pixels in each direction (left, right, top, or bottom); the resulting region is down-sized or up-sized by down-sampling or up-sampling to fit into a 64$\times$64 image, as shown here, defined as an ROI, and is then fed into a DNN for inference.}
\label{fig3}
\end{figure}

Two methods were followed to pre-process APA-frames in preparation for DNN classification; classification was carried out with a VGG16b network \cite{vgg16} trained and tested independently for each method on a GPU:
\begin{itemize}
\item Method 1: In the first method, noise removal was minimally applied to each APA-frame, and the resulting image was re-sized by down-sampling it into a $600\times600$ image, to be used for DNN inference.
\item Method 2: In the second method, aggressive noise removal was applied to each APA-frame before down-sizing the image for inference, followed by cropping around a signal ``region of interest'' (ROI), and re-sizing the resulting ROI (by down-sampling or up-sampling) into a 64$\times$64 image. The noise removal and ROI finding were informed by studying the ADC distributions of simulated APA-frames of different signatures, as shown in Fig.~\ref{fig2}. Examples of ROIs are shown in Fig.~\ref{fig3}. 
\end{itemize}

For both methods, the images were used to train a customized VGG16b network, and the resulting network was tested on a statistically independent sample of images, prepared in the same way, for accuracy and inference speed. The tests were performed on a single GPU, NVIDIA GeForce GTX 1080 Ti.

Inference results on GPU from each method for VGG16b are summarized in Tabs.~\ref{tab2} and~\ref{tab3}. The tables show the number of ROI images used for training and testing for each sample; and resulting accuracy, identified in terms of the fraction of input images in the testing case which get classified under each label: background noise (NB), low energy supernova neutrino interaction (LE), or high energy interaction (HE). The given fractions are inclusive of all event energies. Finally, per-APA-frame inference times are provided, in milliseconds, and include image input i/o from host (GPU server) memory. The key table parameters are the correct classification rates of low energy and high energy frames, both of which are required to be high by DUNE physics performance requirements, as well as the mis-classification rate of noise frames as low energy frames, which should be as low as possible by data reduction requirement considerations. Both methods are found to yield comparable results in terms of classification accuracy. More specifically, the networks are able to select high energy and low energy frames with efficiencies in excess of 95\% and 90\%, respectively. Required efficiency for high energy frames should be $>99\%$ for interactions with visible energy in excess of 100~MeV. The obtained efficiencies are integrated over all energies (which extend below 100~MeV); it is expected that a HE efficiency calculated relative to interactions with visible energy in excess of 100~MeV would be higher. While signal efficiency performance is comparable for the two methods, Method 2 performs much better with respect to mis-classification rates for background noise frames as LE frames, where a false pass rate of 0.35\% is achieved. 

\begin{table}[t]
\caption{GPU inference results using Method 1, obtained with a with VGG16b network (training for 2 epochs and learning rate set to $2\times10^-4$).}
\begin{center}
\begin{tabular}{|l|c|c|c|c|c|c|}
\hline
 & \textbf{Train} & \textbf{Test} & \multicolumn{3}{c|}{\textbf{Accuracy (\%)}} & \textbf{Inference} \\
\textbf{Sample} & \textbf{Size} & \textbf{Size} & $\epsilon_{NB}$ & $\epsilon_{LE}$ & $\epsilon_{HE}$ & \textbf{Time (ms)} \\
\hline
NB & 51,100 & 99,000 & 91.45 & 8.49 & 0.06 & \\
LE & 44,900 & 29,800 & 3.17 & 96.83 & 0 & 27.7$\pm$8.6\\
HE & 52,828 & 67,178 & 6.03 & 3.48 & 90.48 & \\
\hline
\end{tabular}
\label{tab2}
\end{center}
\end{table}

\begin{table}[t]
\caption{GPU inference results using Method 2, obtained with the VGG16b network (training for 13 epochs and learning rate set to 10$^-4$). NB$^*$ corresponds to explicitly non-empty background noise ROIs, containing noise artifacts, which represent approximately 2\% of the regions found after noise removal.}
\begin{center}
\begin{tabular}{|l|c|c|c|c|c|c|}
\hline
 & \textbf{Train} & \textbf{Test} & \multicolumn{3}{c|}{\textbf{Accuracy (\%)}} & \textbf{Inference} \\
\textbf{Sample} & \textbf{Size} & \textbf{Size} & $\epsilon_{NB}$ & $\epsilon_{LE}$ & $\epsilon_{HE}$ & \textbf{Time (ms)} \\
\hline
NB & 12,023& 4,027 & 99.65 & 0.35 & 0 & \\
NB$^*$ & 12,023& 293 & 79.9 & 19.8 & 0.34 & 5.0$\pm$0.3\\
LE & 12,050& 3,970 & 3.78 & 95.04 & 1.18 & \\
HE & 10,137& 3,417 & 2.99 & 6.88 & 90.14 & \\
\hline
\end{tabular}
\label{tab3}
\end{center}
\end{table}

Inference latency for the two methods is also comparable, although Method 2 inference is faster by more than a factor of five, due to the reduced size of the input image. Latency considerations determine whether frame-by-frame inference can be applied during the low-level data selection stage of the DUNE far detector DAQ system; such application would have to keep up with the frame rate of 66.6$\times$10$^3$~fsps. In the case of Method 1, if we required that every frame go through image classification, the observed latency of 27.7~ms (an order of magnitude off 2.25~ms even with a 150-fold parallelization) would preclude such application during low-level data selection, unless a more-than-10-fold parallelization of frame-by-frame processing were to be implemented; application at high-level filter stage, however, is viable, because a relatively low module-level trigger rate (for example of order 1~Hz readout of 200 APA-frames) would make data rate handling more manageable. In the case of Method 2, the inference latency (comparable to APA-frame length of 2.25~ms) is far more promising for a frame-by-frame online low-level data selection implementation; furthermore, the processing time requirement for this method can be relaxed further based on the additional reduction of frame rate gained by the aggressive noise removal and ROI formation pre-processing stage. We have found that after noise removal and ROI finding, only 2\% of the 2.25~ms-long background noise frames survive. Considering that most APA-frames that DUNE will be reading out will contain only background noise, we expect that the average frame rate reduction factor gained will be close to that of the background noise reduction factor. Hence, directing only ROIs containing non-zero pixels to network inference, for example, could relax the processing time requirement by a factor of 50.

\begin{table}[t]
\caption{GPU inference results using Method 2, obtained with the CNN\_s network (training for 48 epochs and learning rate set to $2\times10^-3$).}
\begin{center}
\begin{tabular}{|l|c|c|c|c|c|c|}
\hline
 & \textbf{Train} & \textbf{Test} & \multicolumn{3}{c|}{\textbf{Accuracy (\%)}} & \textbf{Inference} \\
\textbf{Sample} & \textbf{Size} & \textbf{Size} & $\epsilon_{NB}$ & $\epsilon_{LE}$ & $\epsilon_{HE}$ & \textbf{Time (ms)} \\
\hline
NB &12,023 & 4,027 & 99.53 & 0.47 & 0.12 & \\
LE &12,050 & 3,970 & 4.01 & 94.48 & 1.51 &1.6$\pm$0.1 \\
HE &10,137 & 3,417 & 3.63 & 6.15 & 90.22 & \\
\hline
\end{tabular}
\label{tab5}
\end{center}
\end{table}

Additional fake (background noise) trigger reduction is possible at the module-level data selection stage, by aggregating APA-frames classified as LE and considering their coincidence over the anticipated duration (10 seconds) of a supernova burst, following the methodology for supernova burst triggering in \cite{dunetdr}. Findings from preliminary studies \cite{dunetdr,mltn} support the successful application of the coincidence-based methodology fed by CNN-based (using a VGG16b network) low-level information. % resting on the assumption that simulations can be developed to sufficiently accurately predict detector raw data.  

The promise of Method 2 for online application for low-level data selection further motivates the use of smaller networks, and, for the purposes of further acceleration on FPGA, smaller input images. The second method was therefore further explored for a number of other customizable networks, besides VGG16b \cite{vgg16}, including a smaller, simpler CNN than VGG16b, referred to as CNN\_s \cite{cnn}, a Multi-Layer Perceptron (MLP) network \cite{mlp}, and a ResNet14b network \cite{resnet}. Results from the three additional networks are provided in Tabs.~\ref{tab5} through \ref{tab7}, to be considered in comparison with VGG16b results in Table~\ref{tab3}. The best performance is obtained with VGG16b and CNN\_s. The simple CNN (CNN\_s) performs comparably with VGG16b in terms of the accuracy, albeit with with slightly higher pass rate ($\sim$0.5\%) on background noise ROIs. MLP and ResNet14b also have comparable pass rates ($\sim$0.5\%) for background noise, but the accuracies for low energy and high energy ROIs are not as high as those for VGG16b or CNN\_s. Inference times with CNN\_s (on a single GPU card) are an order of magnitude lower than for VGG16b, due to the reduced number of layers and convolutions per layer. 

%\begin{table}[htbp]
%\caption{Pre-processing results for Method 2.}
%\begin{center}
%\begin{tabular}{|l|c|c|}
%\hline
% & \multicolumn{2}{c|}{\textbf{Fraction of non-empty images (\%)}} \\
%\textbf{Sample} & After noise removal & After ROI finding \\
%\hline
%NB & \textcolor{red}{xx} & \textcolor{red}{xx}\\
%LE & \textcolor{red}{xx} & \textcolor{red}{xx}\\
%HE & \textcolor{red}{xx} & \textcolor{red}{xx}\\
%\hline 
%\end{tabular}
%\label{tab4}
%\end{center}
%\end{table}

\begin{table}[t]
\caption{GPU inference results using Method 2, obtained with the MLP\_1 network (training for 65 epochs and learning rate set to $2\times10^-4$).}
\begin{center}
\begin{tabular}{|l|c|c|c|c|c|c|}
\hline
 & \textbf{Train} & \textbf{Test} & \multicolumn{3}{c|}{\textbf{Accuracy (\%)}} & \textbf{Inference} \\
\textbf{Sample} & \textbf{Size} & \textbf{Size} & $\epsilon_{NB}$ & $\epsilon_{LE}$ & $\epsilon_{HE}$ & \textbf{Time (ms)} \\
\hline
NB & 12,023& 4,027 & 99.50 & 0.45 & 0.05 & \\
LE & 12,050& 3,970 & 4.48 & 89.70 & 5.82 & 1.0$\pm$0.08\\
HE & 10,137& 3,417 & 7.29 & 13.08 & 79.63 & \\
\hline
\end{tabular}
\label{tab6}
\end{center}
\end{table}

\begin{table}[t]
\caption{GPU inference results using Method 2, obtained with the ResNet50 network (training for 30 epochs and learning rate set to $10^-5$).}
\begin{center}
\begin{tabular}{|l|c|c|c|c|c|c|}
\hline
 & \textbf{Train} & \textbf{Test} & \multicolumn{3}{c|}{\textbf{Accuracy (\%)}} & \textbf{Inference} \\
\textbf{Sample} & \textbf{Size} & \textbf{Size} & $\epsilon_{NB}$ & $\epsilon_{LE}$ & $\epsilon_{HE}$ & \textbf{Time (ms)} \\
\hline
NB & 12,023& 4,027 & 99.28 & 0.55 & 0.17 & \\
LE & 12,050& 3,970 & 3.55 & 88.89 & 7.56 & 15.3$\pm$1.2\\
HE & 10,137& 3,417 & 2.84 & 15.13 & 82.03 & \\
\hline
\end{tabular}
\label{tab7}
\end{center}
\end{table}

Finally, we note that lower background noise pass rates could be achievable using a variation of a CNN-based selection. For example, in \cite{mltn}, Method 1 is used to train against six classes: NB, LE, plus the four subclasses of the HE class including atmospheric neutrino interactions (atm), nucleon decay (ndk), neutron antineutron oscillation (nnbar), and cosmic interactions (cosmic). Rather than classifying frames in terms of the six labels according to the label returning the highest score, a cut on the NB classification score is applied in order to reject frames with high enough NB scores, and select all surviving APA-frames. Results based on this classification scheme are summarized in Tab.~\ref{tab8} as a function of NB score cut. The number of ROI images used for training and testing for each sample in Tab.~\ref{tab8} correspond to those given in Tab.~\ref{tab2}. The main difference relative to Tab.~\ref{tab2} is that accuracy is identified in terms of the fraction of input images in the testing case which have NB score lower than what is indicated on the left column. (Here, too, fractions are inclusive of all energies.) The average inference time is comparable to that presented in Tab.~\ref{tab2}, and includes image input i/o from host (GPU server) memory.   

\begin{table*}[t]
\caption{GPU inference results using Method 1, obtained with the VGG16b network (training for 2 epochs and learning rate set to $2\times10^-4$), trained on six class labels. See text for more details.}
\begin{center}
\begin{tabular}{|l|c|c|c|c|c|c|c|}
\hline
  & \multicolumn{7}{c|}{\textbf{Accuracy (\%)}} \\
\textbf{NB cut} & $\epsilon_{NB}$ & $\epsilon_{LE}$ & $\epsilon_{HE}$ & $\epsilon_{HE:nnbar}$ & $\epsilon_{HE:ndk}$ & $\epsilon_{HE:atm}$ & $\epsilon_{HE:cosmic}$ \\
\hline
 0.1 &  0.73 & 88.18 & 96.12 & 99.98 & 99.29 & 92.24 & 92.57 \\
 0.01 &  0.14 & 83.27 & 95.68 & 99.98 & 99.18 & 91.01 & 92.46 \\
 0.001 &  0.033 & 77.11 & 95.21 & 99.98 & 99.05 & 89.76 & 92.23 \\
 0.0001 &  0.011 & 69.74 & 94.61 & 99.97 & 98.74 & 88.39 & 91.71 \\
 0.00001 &  0.002 & 60.73 & 93.79 & 99.95 & 98.22 & 86.61 & 90.97 \\
\hline
\end{tabular}
\label{tab8}
\end{center}
\end{table*}

\section{CNN Implementation in FPGA}
\label{sec4}

The accuracy performance of CNN\_s obtained with reduced-size raw data images, combined with the reduced size of the network relative to VGG16b, motivate studies for further hardware acceleration.  Hardware accelerators can be designed according to two main different approaches~\cite{coupling}: the designer can tightly couple the hardware functional unit inside the pipeline of a processor core or choose a loose out-of-core coupling architecture. Loosely-coupled accelerators (LCA) are hardware accelerators capable of performing Direct-Memory Access (DMA) to external main memory. LCAs are located outside the processor cores, for example on the FPGA fabric, and interact with the rest of the chip through on-chip interconnects. They can implement coarse-grain operations with dedicated datapaths that can accelerate a complete application functionality (e.g. the convolutional layers in the case of a CNN). To implement our accelerators we adopted the LCA approach, as it represents a perfect match in terms of reconfigurability and flexibility with FPGAs and embedded SoCs.

The bottleneck for the performance of inference of CNNs are the convolutional layers, which alone are responsible for more than 90\% of the computations performed on networks like VGG16b. Thus we chose to specifically design a {\it convolutional} LCA for our CNNs. 

While the workloads of many accelerators described in literature are fixed and known at design time~\cite{dse}, a convolutional layer has a number of parameters that are known only at run time (input dimensions, number of input channels, number of filters per layer, etc.). Different configurations of these hyperparameters lead to drastic changes in memory requirements and computational capabilities. Thus, we chose to design a LCA that is {\it configurable} at run-time.

We used High-Level Synthesis (HLS) to obtain the FPGA implementation starting from specifications made in C/C++~\cite{nane}. Current HLS tools enable an effective exploration of the design space of an accelerator to obtain many alternative implementations which are trade-offs of resource/power requirements and performance~\cite{simoneth,prost,liu}.

\subsection{Accelerator Architecture}

\begin{figure}[t]
\centering
\includegraphics[width=0.48\textwidth,trim=0 0 0 0, clip]{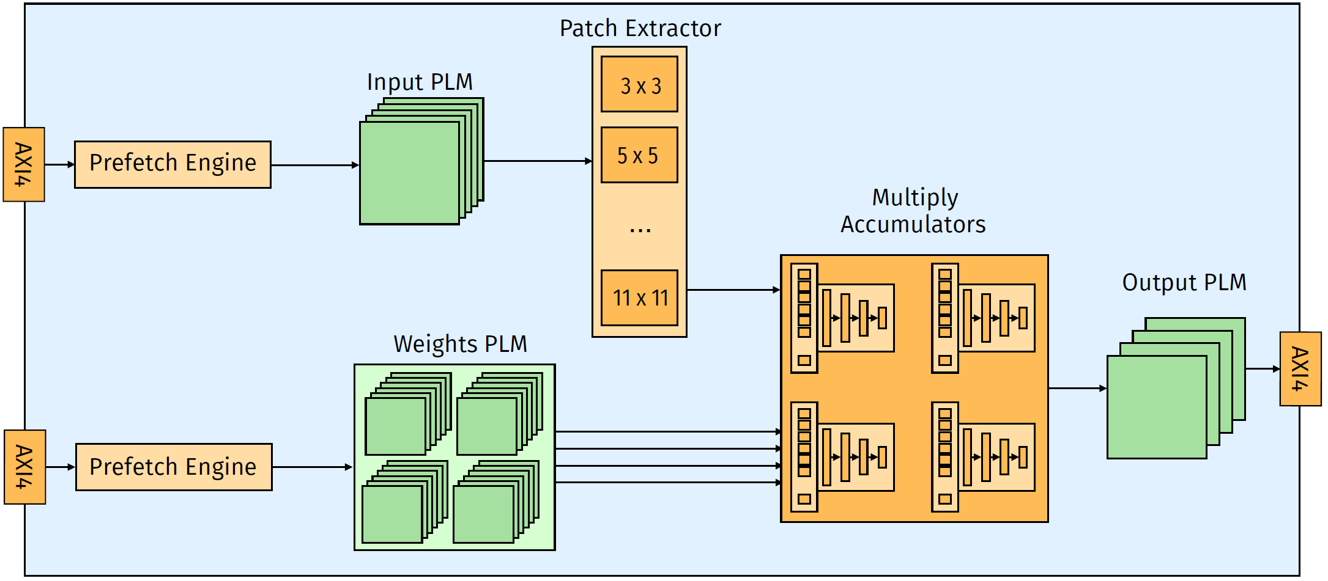}
\caption{Overview of the configurable loosely-coupled accelerator.}
\label{fig4}
\end{figure}

Figure~\ref{fig4} illustrates the main components and memories of our configurable convolutional LCA. It embeds three private local memories (for storing the input and output features and the filter weights), a patch extractor (for data reordering), and several multiply-and-accumulate engines which are the core of the convolution operations. The accelerator communicates with the rest of the chip through AXI4 interconnects~\cite{amba}.

{\it Private Local Memory}. Custom hardware accelerators allow designers to tune the microarchitecture and enable higher level of optimization to meet a specific configuration and workload, providing high performance and energy efficiency~\cite{uproc}. 
General purpose processors (CPUs) leverage the hierarchy of caches and memory to provide the best solution in terms of  bandwidth and latency across a variety of applications. Similarly, GPUs offer very high bandwidth and massive availability of parallel computational cores (CUDA cores for GPUs NVIDIA). When implementing custom hardware accelerator on FPGAs, resource utilization and allocation is an important design constraint. The designer should carefully optimize the accelerator to reuse data as much as possible, thus balancing communication versus computation and reducing expensive memory transfers from the off-chip main memory. This requires the use of private local memories (PLMs), which offer low latency, high bandwidth memory and customizable word widths. They do so, by providing many banks and ports that are exclusively accessed by the datapath logic of the LCA that embeds them~\cite{coupling}. Careful design and tailoring of these structures for input/output ports, partitioning, and resource allocation is essential to constantly provide data to be fed to all the high-performance computational engines.

{\it Patch Extractor}. The patch extractor is an optimized module for retrieving the portion of the input features where the filters are applied. This operation is highly dependent on the choice of hyperparameters. Due to the irregular access pattern that this module performs while fetching data from the Input PLM, we decided to have various implementations for the most common cases, from the smallest 3$\times$3 filters up to bigger 11$\times$11 filters. At run time, accordingly to the settings of the convolutional layer, the accelerator would choose and enable the correct patch extractor.

{\it Multiply-and-Accumulate (MAC)}. The computational core of convolutional layers lies in the MAC operation. The amount of MAC per input image added up quickly from few thousands for LeNet network~\cite{lenet} up to tens billions for VGG16 network~\cite{vgg16}. To meet this computation requirement, our accelerator embeds several MAC engines. Each of these works on an independent input filter, allowing the parallelization of the computation of the output activation map across multiple filters. Internally, each MAC is implemented with a set of multipliers and accumulators. Changing the number of those components directly affects the degree of parallelism.

\subsection{Performance and Power Analysis}

\begin{figure}[t]
\centering
\includegraphics[width=0.3\textwidth,trim=0 0 0 0, clip]{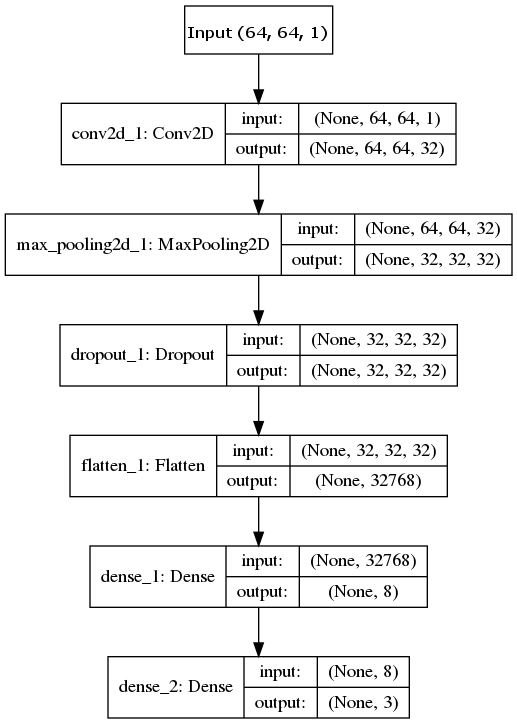}
\caption{Overview of our customized CNN, CNN\_s.}
\label{fig5}
\end{figure}

We ran our tests on a Xilinx Embedded FPGA (Zynq UltraScale+ XCZU9EG MPSoC) that combines both an ARM Cortex-A53 64 bits multi-core processor and FPGA fabric fabricated in 16~nm technology. Overall, it represents a state-of-the-art {\it embedded} platform for a fair evaluation between FPGA acceleration of deep-learning inference tasks and pure software execution. We implemented a customized CNN, CNN\_s ({\it DUNE-CNN-01}), in C language as a reference for our performance and power analysis. Figure~\ref{fig5} provides an overview of CNN\_s.

Table~\ref{tab10} summarizes the results. The inference time of our customized CNN\_s for a single image is 0.0855~seconds when executed as software on the ARM Cortex-A53 CPU. The CPU runs at 1.2~GHz. The inference time of the same network when it leverages the FPGA-acceleration is 0.0511 seconds. The accelerator runs at 100MHz on the FPGA fabric. The total power for the processing system (CPU) and for the FPGA accelerator are 2.871 Watts and 1.110 Watts respectively, as reported in Vivado Power Analysis. The energy efficiency of the FPGA implementation is more than 4 times better than the embedded CPU.

\begin{table}[t]
\caption{Performance and power analysis results on the Embedded FPGA (Zynq UltraScale+ XCZU9EG MPSoC).}
\begin{center}
\begin{tabular}{|l|c|c|c|c|}
\hline
 %& \textbf{Train} & \textbf{Test} & \multicolumn{3}{c|}{\textbf{Accuracy (\%)}} & \textbf{Inference} \\
\textbf{Platform} & \textbf{Model} & \textbf{Time} & \textbf{Power} & \textbf{Energy Efficiency} \\
 & & (s) & (W) & (img/s/W) \\
 \hline
 \textbf{ARM C-A53} & CNN\_s & 0.0855 & 2.871 & 4.074 \\
 \hline
 \textbf{FPGA} & CNN\_s & 0.0511 & 1.110 & 17.630  \\
%\hline
%NB & \textcolor{red}{xx}& 4,027 & 99.28\% & 0.55\% & 0.17\% & \textcolor{red}{xx}\\
%LE & \textcolor{red}{xx}& 3,970 & 3.55\% & 88.89\% & 7.56\% & \textcolor{red}{xx}\\
%HE & \textcolor{red}{xx}& 3,417 & 2.84\% & 15.13\% & 82.03\% & \textcolor{red}{xx}\\
\hline
\end{tabular}
\label{tab10}
\end{center}
\end{table}

% FIX32
%
% Vivado HLS
% BRAM_18K 680   [1824] 37%
% DSP48E   102   [2520]  4%
% FF     14332 [548160]  2%
% LUT    30237 [274080] 11%  
%
% hw/sw O0,g 7.06x / O3 -1.38x

%\textcolor{red}{Giuseppe: Here, describe hardware accelerator design for CNN.}

%Show fpga performance vs. arm CPU for optimized implementation: for each of the two networks and two methods? (vgg16 and resnet14, plus down-sampling and ROI cropping).

\section{Viability of DNN Application for DUNE Data Selection}
\label{sec5}

%xx Georgia: Discussion
%
%Here, describe scalability for GPU and FPGA implementations for DUNE. Identify factor improvements needed to keep within constraints (power, latency, hardware resources), and avenues for further development.
Our studies demonstrate that DNNs in general can meet trigger efficiency requirements for selecting off-beam rare events in the DUNE far detector. In addition, for several CNNs (e.g.~VGG16b), sufficiently low fake trigger rates can be met, such that the required data reduction factor of 10$^4$ can be achieved for high energy triggering and potentially also for low energy triggering with a subsequent module-level data selection stage; the latter is the subject of future investigations.  

For the case of an online data selection implementation where inference is carried out exclusively in GPUs, out of the four DNNs considered, CNN\_s is identified as the most viable option for GPU deployment at the low-level data selection stage. We assume that the necessary pre-processing from preparing the ROIs, which consists of operations which are commonly done in FPGA, can keep up with the raw detector APA-frame rate, and consider only the inference stage latency for the purposes of this discussion. Given that the inference time for an ROI with CNN\_s is comparable to the APA-frame length (2.25~ms), CNN\_s should on average keep up with frame-by-frame selection, with each APA's frames processed in a separate GPU card; this, however, implies that a 200-fold parallelization would be needed (across 200 GPU cards) to facilitate low-level data selection for a 10~kton module; this is unfeasible given power restrictions underground at the far detector location. On the other hand, a factor of 50 reduction in required GPU processing would be possible if a pre-processing step were to be added to remove empty ROIs before the inference stage. Such a step would remove all but 2\% of the background noise ROIs from the inference stage, allowing for, on average, 112.5~ms per ROI for inference. The same scheme would make VGG16b viable for online inference no GPU as well, which yields characteristically higher efficiency for all rare events of interest.

In the case of FPGA inference we find that a factor of four (4) increase in energy efficiency (img/s/W) is possible over a software implementation in CPU of the same (CNN\_s) algorithm, motivating consideration of deployment of CNNs for low-level data selection on FPGA. The performance improvement over a software implementation is comparable for both inference speed (factor of 1.7) and power efficiency (factor of 2.6). Furthermore, we find that for smaller networks, such as for CNN\_s, the resource allocation requirements for a full network implementation processing ROIs of 64$\times$64 size are comparable with those available in state-of-the-art FPGAs, a desirable feature for simplified parallelization and for minimizing costs.

%Comment on xx
%resting on the assumption that simulations can be developed to sufficiently accurately predict detector raw data.  

%\begin{table}[htbp]
%\caption{Table Type Styles}
%\begin{center}
%\begin{tabular}{|c|c|c|c|}
%\hline
%\textbf{Table}&\multicolumn{3}{|c|}{\textbf{Table Column Head}} \\
%\cline{2-4} 
%\textbf{Head} & \textbf{\textit{Table column subhead}}& %\textbf{\textit{Subhead}}& \textbf{\textit{Subhead}} \\
%\hline
%copy& More table copy$^{\mathrm{a}}$& &  \\
%\hline
%\multicolumn{4}{l}{$^{\mathrm{a}}$Sample of a Table footnote.}
%\end{tabular}
%\label{tab1}
%\end{center}
%\end{table}

\section{Summary}
\label{sec6}

Acceleration of DNNs for real-time data selection is motivated by a number of up and coming high-resolution imaging particle detectors, in particular LArTPCs which work by imaging particle traces that are identifiable by their distinct topologies (spatial extent, shape, and pixel intensity) in two-dimensional view projections of three-dimensional detector regions. We have investigated the viability of DNN application for the purposes of real-time or online data selection (triggering) for such detectors, with a particular focus on the future DUNE experiment. Data selection is achieved by frame-by-frame classification of raw data streamed in channel vs.~time space from 200 independent, self-contained regions of one of four DUNE far detector modules, assuming a single phase design. 

Using simulated DUNE raw data images (APA-frames), we have found that such techniques yield promising results in terms of image classification accuracy, for a large variety (in terms of depth and size) of networks. Sufficiently high trigger efficiencies are achieved for selection of APA-frames with high energy interactions; lower trigger efficiencies are achieved for APA-frames with low energy interactions. However, supernova burst trigger efficiency can be optimized further by exploiting a higher-level decision which aggregates selected APA-frames over time, following the approach in \cite{dunetdr}.

We have further shown that latency and power considerations make the implementation of DNNs on GPUs for online inference viable for smaller networks and with significantly re-sized and down-selected ROI image inputs. Larger networks with re-sized full-frame information are viable only for the high level filter stage, at this time. 

Finally, we have shown that implementation of DNNs on FPGAs for real-time inference at the low-level stage is promising, and have provided a viable path for development and optimization.

\section*{Acknowledgment}
%G.K.~and Y.J.~thank J.~Hewes, Y.~Zhou, and S.~Koo for early contributions to the development of CNNs and simulation tools used in the GPU studies. L.C.~and G.D.G.~thank S.~Rossi for early contributions to the development of \textcolor{red}{xx}. All authors would like to thank K.~Terao for valuable input and feedback to this work.
The authors thank J.~Hewes, Y.~Zhou, and S.~Koo for early contributions to the development of CNNs and simulation tools used in the GPU studies, S.~Rossi for a preliminary analysis of the FPGA implementation of the CNN, and K.~Terao for valuable input and feedback to this work. This material is based upon work supported by the National Science Foundation under Grant No.~PHY-1753228, and work supported in part by the Research Initiatives in Science and Engineering (RISE) program of Columbia University.

%\section*{References}

%Please number citations consecutively within brackets \cite{b1}. The 
%sentence punctuation follows the bracket \cite{b2}. Refer simply to the reference 
%number, as in \cite{b3}---do not use ``Ref. \cite{b3}'' or ``reference \cite{b3}'' except at 
%the beginning of a sentence: ``Reference \cite{b3} was the first $\ldots$''

%Number footnotes separately in superscripts. Place the actual footnote at the bottom of the column in which it was cited. Do not put footnotes in the abstract or reference list. Use letters for table footnotes.

%Unless there are six authors or more give all authors' names; do not use ``et al.''. Papers that have not been published, even if they have been submitted for publication, should be cited as ``unpublished'' \cite{b4}. Papers that have been accepted for publication should be cited as ``in press'' \cite{b5}. Capitalize only the first word in a paper title, except for proper nouns and element symbols.

%For papers published in translation journals, please give the English citation first, followed by the original foreign-language citation \cite{b6}.

%\vspace{12pt}
%\color{red}
%IEEE conference templates contain guidance text for composing and formatting conference papers. Please ensure that all template text is removed from your conference paper prior to submission to the conference. Failure to remove the template text from your paper may result in your paper not being published.

\end{document}